\documentclass[letter]{aa}
\usepackage{graphicx}
\usepackage[varg]{txfonts}

\def\Msun{M_\odot}
\def\lsim{\mathrel{\rlap{\lower 3pt \hbox{$\sim$}} \raise 2.0pt \hbox{$<$}}}
\def\gsim{\mathrel{\rlap{\lower 3pt \hbox{$\sim$}} \raise 2.0pt \hbox{$>$}}}
\newcommand{\frc}[2]{\tiny{\raisebox{2pt}{$#1$}\big/\raisebox{-3pt}{$#2$}}}

\begin{document}

\title{Spatial regularity of the young stellar population in the ring \\
       of NGC~6217}

\author{Alexander~S.~Gusev
          \thanks{e-mail: gusev@sai.msu.ru}
          \and Elena~V.~Shimanovskaya
          }

\institute{Sternberg Astronomical Institute, Lomonosov Moscow State University, 
           Universitetsky pr. 13, 119234 Moscow, Russia}

\titlerunning{Regularity of the young stellar population in the ring}
\authorrunning{Gusev and Shimanovskaya}
           
\date{Received May 29, 2020; Accepted July 15, 2020}

\abstract{The relative contribution of various physical processes to the 
spatial and temporal distribution of molecular clouds and star-forming 
regions in the disks of galaxies has not yet been the subject of extensive study. 
Investigating the spatial 
regularity in the distribution of the young stellar population in spiral 
and ring structures is a good test for studying this contribution. In this paper, we 
look at the photometric properties of the ring and spiral arms in the barred 
spiral galaxy NGC~6217 based on an analysis using {\it GALEX} ultraviolet, optical 
$UBVRI,$ and H$\alpha$ surface photometry data. The ring in the galaxy 
is located near the corotation area. We found evidence of spatial regularity in the 
distribution of the young stellar population along the galaxy ring. The 
characteristic scale of spacing is about 700~pc. At the same 
time, we did not find a similar regularity in the distribution of the young 
stellar population along the spiral arms of NGC~6217. The spatial regularity 
in the concentration of young stellar groupings along spiral arms is a quite 
rare phenomenon and it has never previously been seen in galactic rings.}

\keywords{Galaxies: individual: NGC~6217 -- Galaxies: star formation -- 
          Galaxies: ISM}

\maketitle

\section{Introduction}

Various physical processes, such as gravitational collapse and turbulence 
compression, are believed to play a main role in the creation of molecular 
clouds and their further evolution into star formation regions. The sizes of 
these structures and their spatial distribution in galactic disks can be 
explained in terms of gravitational or magnetogravitational instability 
\citep[see, e.g.,][]{elmegreen1983,elmegreen1994b}. However, the influence 
of the magnetic field and shock waves on the spatial distribution of gas clouds 
and young stellar groups is still not fully understood. Galactic spiral arms 
and rings are good laboratories for studying the relative effect of different 
physical processes on the spatial distribution of star-forming regions.

The regular spatial distribution of young star complexes is a rather 
unique phenomenon which was visually detected by \citet{elmegreen1983} 
in spiral arms of only 22 grand design galaxies. We note that among these 
galaxies, present in only seven stellar systems, the regular chains of star 
complexes are observed in both spiral arms. The data of 
\citet{elmegreen1983} remains the only published result of a systematic 
search for regular chains of stellar complexes in spiral arms.

In later years, \citet{efremov2009,efremov2010,efremov2011} found the regularities 
in the distribution of H\,{\sc ii} superclouds in spiral arms of our 
Galaxy and in the distribution of star complexes in M31. \citet{gusev2013} 
confirmed the result of \citet{elmegreen1983}, finding that only one of the spiral 
arms in NGC~628 has a regular chain of star complexes with a separation of 
$\sim1.7$~kpc. However, they also found the existence of the same characteristic 
separation ($\approx400$~pc) between adjacent fainter star-formation regions 
in both spiral arms of the galaxy. Finally, \citet{elmegreen2018} found 
regularly spaced infrared peaks in the dusty spiral arms of M100 with a 
typical spacing between the clumps of $\sim410$~pc.

\begin{table}
\caption{Basic parameters of NGC~6217.}
\label{table:galgen}
\centering
\begin{tabular}{ll}
\hline \hline
Parameter                                & Value \\
\hline
Type                                     & (R'L)SB(rs)b \\
Total apparent $B$ magnitude ($B_t$)     & 11.89 mag \\
Absolute $B$ magnitude ($M_B$)$^a$       & $-20.45$ mag \\
Inclination ($i$)                        & $33\degr$ \\
Position angle (PA)                    & $162\degr$ \\
Apparent corrected radius ($D_{25}$)$^b$ & 2.30 arcmin \\
Apparent corrected radius ($D_{25}$)$^b$ & 13.8 kpc \\
Distance ($d$)                           & 20.6 Mpc \\
Galactic absorption ($A(B)_{\rm Gal}$)   & 0.158 mag \\
\hline
\end{tabular}
\tablefoot{
\tablefoottext{a}{Absolute magnitude of a galaxy corrected for 
Galactic extinction and inclination effect.}
\tablefoottext{b}{Isophotal diameter (25 mag\,arcsec$^{-2}$ in 
the $B$-band) corrected for Galactic extinction and absorption 
due to the inclination of NGC~6217.}
}
\end{table}

\begin{figure}
\centering
\includegraphics[width=8.9cm]{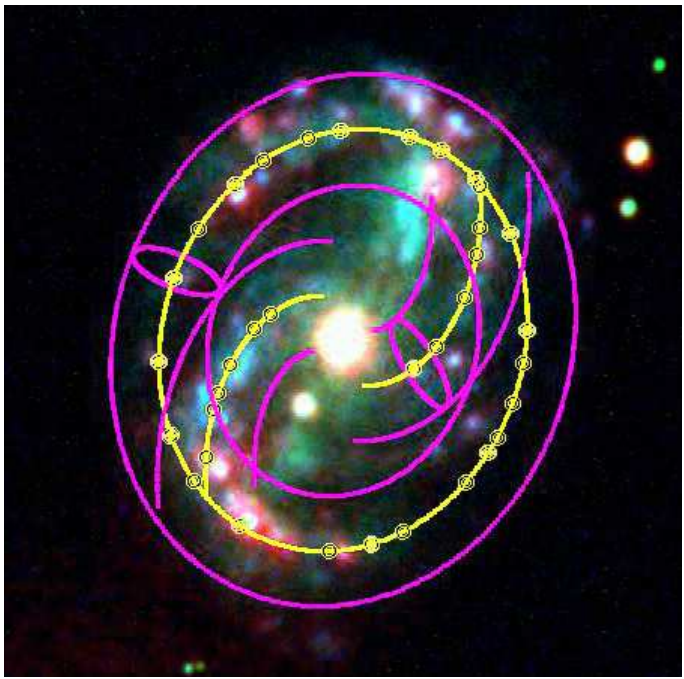}
\caption{Composite image of NGC~6217 in the $U$ (blue), $B$ (green) passbands, 
and H$\alpha$ line (red) with the logarithmic spiral segments and the ring 
(yellow lines) overlaid. The yellow-black-yellow circles on the curves of the arm and ring 
correspond to the projected positions of the local maxima of brightness. 
The areas of flux measurements in apertures are bounded by magenta ellipses 
(segments). Examples of elliptical apertures are shown in magenta. The 
size of the image is $120\times120$ arcsec$^2$. North is upward and east is 
to the left.}
\label{fig:map}
\end{figure}

Along with spiral arms, galactic rings are the product of evolutionary 
processes taking place in galactic disks. These appear to be related to specific orbital 
resonances with the pattern speed of a bar \citep{buta2017}. Unlike 
arms, the physical and dynamic parameters of medium (gas density, temperature, 
pressure, rotation velocity) along rings are likely to be constant. On the 
other hand, strong shock waves are not expected to be present in rings. These 
features make them interesting objects for the study of the spatial distribution 
of star-forming regions.

Based on general ideas about physics and evolution of rings, investigating spatial 
regularity in the distribution of star clusters, as well as in arms, 
is not an impossible undertaking. \citet{elmegreen1994a} theoretically described the 
possibility of the existence of regularly spaced starburst spots in galactic 
rings. However, this had not been seen in observations before. Only \citet{artamonov1999} 
noted quasi-periodic characteristic distances of $1.3\pm0.2$~kpc between 
adjacent star complexes in the ringed barred galaxy NGC~6217.
The purpose of our study is to verify the possible regularity in the 
distribution of star formation regions along the ring of NGC~6217.

NGC 6217 is a galaxy with a prominent bar, two strong spiral arms, and two 
rings. One peculiarity of the galaxy is that its spiral arms begin from the 
nuclear region and, upon crossing the ends of the bar, they form a (pseudo-)ring 
with a major projected radius of 40.2 arcsec \citep{comeron2014} 
(Fig.~\ref{fig:map}). The faint outer  ring has radius that is $\approx2$ times 
larger then the inner one \citep[81.6 arcsec according to][]{comeron2014}. However, 
  it is not considered in this paper and so, it is not shown in Fig. 1.

The photometry and kinematics of the galaxy have been studied in a wide range 
of wavelengths from UV to radio \citep[see][and references therein]
{driel1991,elmegreen1996,artamonov1999,james2004,hernandez2005,garrido2005,
font2019}. We note that the NGC~6217 inner ring that is the subject of current studies is 
located near the corotation region \citep{driel1991,elmegreen1996,font2014}. 

The fundamental parameters of NGC~6217 are presented in
Table~\ref{table:galgen}, where the morphological type is taken from 
\citet{buta2015}, Galactic absorption, $A(B)_{\rm Gal}$, taken from the 
NED\footnote{http://ned.ipac.caltech.edu/} database, and the remaining 
parameters are taken from the LEDA\footnote{http://leda.univ-lyon1.fr/} data 
base. The adopted value of the Hubble constant is equal to 
$H_0 = 75$~km\,s$^{-1}$Mpc$^{-1}$. With the assumed distance to NGC~6217, we 
estimate a linear scale of 100~pc\,arcsec$^{-1}$. We note that the kinematic 
positional angle of the galaxy ($287\degr$) is significantly different from 
the photometric one \citep{font2019}. However, given the low inclination of 
NGC~6217, this does not substantially affect our results.

\section{Observational data}

The results of the $UBVRI$ photometry for the galaxy were published 
in \citet{gusev2015}. We also use the H$\alpha$ image, obtained in the {\it GHASP} 
survey \citep{epinat2008} and downloaded from the NED data base, in our study. The absolute 
calibration of this image was carried out using the parameters of descriptors 
from the original FITS file (image) and their explanations in 
\citet{epinat2008}. We already used H$\alpha$ data for NGC~6217 in 
\citet{gusev2018}.

Ultraviolet {\it GALEX} NUV-reduced FITS-images of NGC~6217 were 
downloaded from the B.~A.~Miculski Archive for space 
telescopes.\footnote{http://galex.stsci.edu/GR6/; 
files AIS\_21\_sg76-nd-*.fits} The observations 
were carried out on November 11, 2011, with a total exposure of 96~s. 
The description of the {\it GALEX} mission, parameters of NUV passband, 
files description, and data reduction are presented in \citet{morrissey2005}.

Images of the galaxy in all bands were transformed to the face-on position 
using the values of $i$ and the position angle (PA) from Table~\ref{table:galgen}. Those 
face-on images were used for the further analysis. Photometric 
magnitudes are corrected for Galactic extinction and for projection effects: 
$m_{\rm cor} = m_{\rm obs}\,10^{0.4A_{\rm Gal}}\cos(i)$.
Our image resolution is equal to 1.7 arcsec for $U$, 1.4 arcsec for $B$ and 
$V$, 3.3 arcsec for H$\alpha$, and 5.3 arcsec for NUV.

\section{Results}

\subsection{Along-ring and -arm photometry}

For the next steps in our analysis, we used a technique developed by us in 
\citet{gusev2013}. In the first step, we define the geometric parameters 
of the spiral arms and the ring by fitting the curves of both main arms and 
the ring, which are clearly outlined in the optical bands (see 
Fig.~\ref{fig:map}). The arms and the ring are defined by a visual selection 
of pixels in the part of these images that belongs to the center lines of the spiral 
arms or the ring. Pixels in this region are then fitted with a logarithmic spiral 
for the arms or a circle for the ring using linear least-squares. Here we assume 
that the pitch angle $\mu$ is constant along the arm and equal for both 
spirals.

Using the classic equation for the logarithmic spiral 
$r = r_0\exp{(k(\theta-\theta_0))}$, where $k = \tan \mu$, we obtained 
the following parameters for the spiral arms, $r_0 = 9.65$ arcsec (960~pc) and 
$k = 0.75\pm0.07$, which corresponds to a pitch angle $\mu = 36.8\pm2.5\degr$. 
The zero point angle $\theta_0 = 30\degr$ for the eastern arm and $210\degr$ for 
the western arm (where $\theta$ is counted from north towards east, the same 
as the PA; see Fig.~\ref{fig:map}). As a result of the fitting, we also 
obtained a radius of the ring $r_{\rm ring} = 38\pm4$ arcsec ($3.8\pm0.4$~kpc). 
The ring parameters used by us ($r$ and PA) are close to those obtained 
by \citet{comeron2014}. The resulting spiral arms and the ring are shown in 
Fig.~\ref{fig:map} in projection to the apparent plane of the galaxy.

We note the extremely high pitch angle, which is $\approx2$ times larger than the 
average value for galaxies of the same morphological type \citep{diaz2019}. 
Apparently, this is a consequence of the high central-mass concentration in 
NGC~6217. It is known that the galaxy has a very bright nucleus in a wide range 
of wavelengths \citep[see][and references therein]{williams2019}, a high 
H\,{\sc i} central surface density \citep{driel1991}, and it exhibits signs of 
nuclear activity \citep{veron2010}.

\begin{figure}
\vspace{3mm}
\centering
\includegraphics[width=8.9cm]{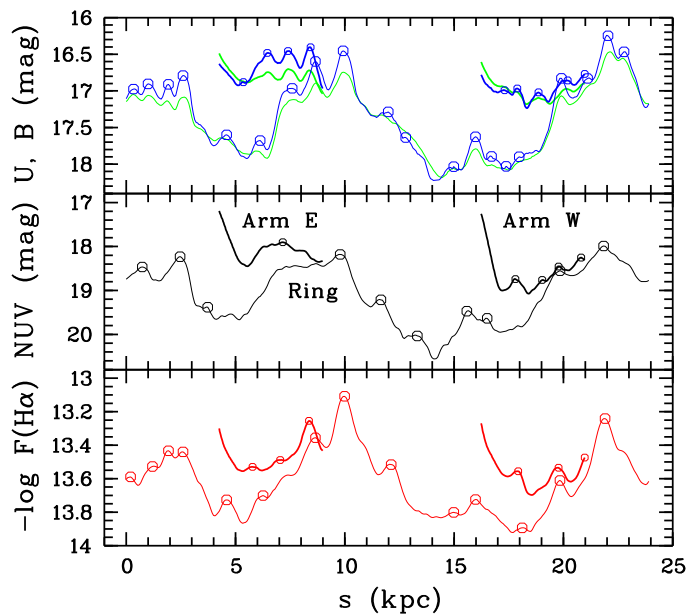}
\caption{Photometric profiles in the $U$ (blue), $B$ (green), NUV (black) 
bands, and H$\alpha$ line (red) along the ring (thin curves) and the 
spiral arms (thick curves). The ordinate units are magnitudes 
and logarithm of H$\alpha$ flux in units of erg\,s$^{-1}$cm$^{-2}$ within 
the aperture. Positions of local maxima of brightness on the profiles 
in the $U$, NUV bands, and H$\alpha$ line are indicated by circles for 
the ring (large symbols) and for the arms (small symbols).}
\label{fig:profile}
\end{figure}

\begin{figure}
\vspace{2mm}
\centering
\includegraphics[width=8.9cm]{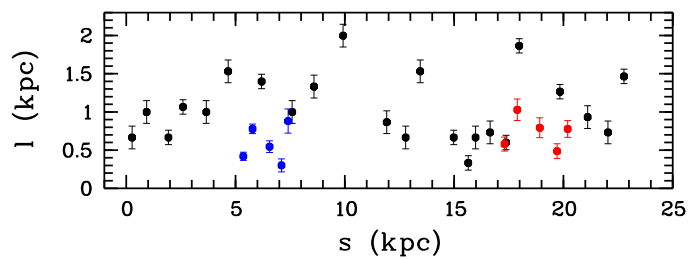}
\caption{Separations $l$ between adjacent local maxima of brightness along 
the ring (black circles), Arm~E (blue), and Arm~W (red). The error 
bars are shown. Positioning errors, $\Delta s$, are smaller than the 
circles size. See the text for details.}
\label{fig:separate}
\end{figure}

To study variations in brightness along the ring and spiral arms, we obtained 
photometric profiles along these structures. We used the elliptical aperture 
($20\times6$ arcsec$^2$), with a minor axis along the spiral arm and ring (see 
Fig.~\ref{fig:map}) with a step of $1\degr$ by PA This step corresponds 
to a linear distance of $\approx67$~pc for the ring and varies from 
$\approx20$ to $\approx80$~pc for the spiral arms. The strip of 20~arcsec 
wide captures all star-forming regions in the ring (spiral arms), but does 
not include areas outside them. The exception is the large star complex in the 
northwestern part of the bar (Fig.~\ref{fig:map}), however, in the analysis it 
is superimposed on the young stellar region in the ring having the same PA.
The obtained photometric profiles in NUV, $B$, $V$, and H$\alpha$ along the 
ring and arms are presented in Fig.~\ref{fig:profile}.

\subsection{Spatial regularity of young stellar population along the ring 
            and arms}

Longitudinal displacement along the ring, denoted as $s$, is equal to 
$r_{\rm ring}(\theta-\theta_0)$, where $\theta$ is in radians. For the 
logarithmic spirals, $s = (\sin \mu)^{-1}r_0(\exp{(k(\theta-\theta_0))}-1)$. 
We adopted a zero-point for the displacement along the ring, $s=0$,  
corresponding to PA$=0$. It increases from north toward east, the same as 
the PA. The displacement values along the spiral arms were combined with the 
displacements along the ring: at the points of intersection between the spiral 
arms and the ring at the outer ends of the arms, their displacement was 
taken equal to the displacement along the ring, $s_{\rm arm}=s_{\rm ring}$.

Using mostly profiles in the $U$ band and H$\alpha$ line, and involving 
profiles in the NUV band, we found the local brightness maxima on the 
profiles. We prefer to use the $U$ band image, since it has sufficient 
resolution and is sensitive to the presence of a young stellar population. 
In addition, the H$\alpha$-image was used to identify H\,{\sc ii} regions 
those are weakly visible in the $U$ passband. The low-resolution NUV image, 
as well as the image in the $B$ band, were used for control.

The local maxima of brightness were determined as the lower extrema of the 
functions $m_U(s)$, $m_{\rm NUV}(s)$, $-\log F({\rm H}\alpha)(s)$ for the 
ring and both spiral arms. To locate them, we looked for points with the 
first derivative of the function, $dm/ds = 0$ or 
$d(\log F({\rm H}\alpha))/ds = 0$, and the second derivative, 
$d^2m_U/ds^2 > 0$ or $d^2(-\log F({\rm H}\alpha))/ds^2 = 0,$ on the 
profiles. We selected peaks whose widths exceed three measurement points 
(corresponding to the angular resolution of the images in optical bands) 
and whose amplitude exceeds 0.043 mag for the $U$, 0.017 dex 
for H$\alpha$ and 0.24 mag for the NUV (which corresponds to the threshold 
$3\sigma$ above the average background level in the corresponding images 
within the used aperture, $20\times6$ arcsec$^2$).

As a result, we obtained 23 local maxima of brightness in the ring, with 
6 maxima in the eastern arm (Arm~E), and 6 maxima in the western arm 
(Arm~W). The positions of these maxima are shown in Fig.~\ref{fig:profile}. 
Some local maxima of brightness do not appear in $U$ or in H$\alpha$ 
(see Fig.~\ref{fig:profile}), which is a reflection of the photometric 
evolution of young star clusters with an age from $\sim3$ to $\sim300$~Myr 
\citep{whitmore2011}.
The typical errors of maximum positions are $\Delta s = \pm1$ measurement 
points, which corresponds to an error of $\pm70$~pc for the ring; the maximum 
$\Delta s$ does not exceed 140~pc (see Fig.~\ref{fig:separate}).

In the next step, we measure separations, $l$, between adjacent local 
maxima of brightness along the ring and arms (Fig.~\ref{fig:separate}). 
The separation between the $n^{\rm th}$ and $(n+1)^{\rm st}$ maxima is defined as 
$l_n = s_{n+1}-s_n$. For the ring, we looped the profile so that 
$l_{23} = s_1-s_{23}$.
Figure~\ref{fig:separate} shows a rather regular separation between 
adjacent local maxima of brightness along the ring. 

\begin{figure}
\vspace{3mm}
\centering
\includegraphics[width=8.9cm]{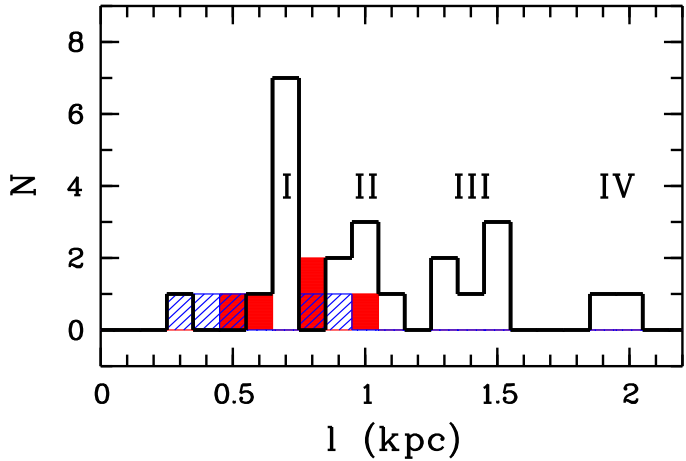}
\caption{The number distribution histograms of local maxima of brightness 
by separation between adjacent objects along the ring (thick black line), 
Arm~E (shaded in blue), and Arm~W (filled red). Roman numerals denote 
subsets of the objects with close characteristic separations.}
\label{fig:hist}
\end{figure}

We built a histogram of the distribution of separations $l$ in the ring 
and in the spiral arms (Fig.~\ref{fig:hist}). We note that the arm 
distributions do not show noticeable regularity. For the ring, the 
histogram  shows a well-pronounced main peak in the distribution at 
$l\approx700$~pc and several secondary peaks. 

We estimated the mean and median separations between local maxima of 
brightness in the ring of the galaxy for four separate subsets of objects, 
numbered in Fig.~\ref{fig:hist}. Our results are presented in 
Table~\ref{table:mean}, where the confidence intervals correspond to 
a 95\% confidence probability (0.05 significance level in the framework of 
Student's $t$-test). The errors in the table do not include the intrinsic 
measurement errors in pairs, that is, equal to $\approx0.1$~kpc, which does not 
depend on the affiliation of pairs to a particular subset.

\begin{table}
\caption{Characteristic separations $l$ of local maxima of brightness 
in the ring.}
\label{table:mean}
\centering
\begin{tabular}{lrcc}
\hline \hline
Subset & $N$ & Mean (kpc) & Median (kpc) \\
\hline
I   & 8 & $0.67\pm0.04$ & $0.67$ \\
II  & 6 & $0.98\pm0.07$ & $1.00$ \\
III & 6 & $1.42\pm0.11$ & $1.40$ \\
IV  & 2 & $1.93\pm0.09$ & $1.86$ \\
\hline
\end{tabular}
\end{table}

We also note that there is no correlation between the affiliation of pairs 
of local maxima of brightness to different subsets and the accuracy of 
measurement of their positions and peak amplitudes.

Thus, the distribution of separations can be multimodal. Given  the 
sample size is too small to make a robust multimodality test, we calculated 
distances $D$ between the peaks relative to their widths according to the 
definition by \citet{ashman1994}: 
$D=\vert \mu_a-\mu_b \vert / \sqrt{(\sigma_a^2+\sigma_b^2)/2}$, where 
$\mu_a$, $\mu_b$ are the means of subsets $a$ and $b$ (I and II or II and 
III), $\sigma_a$, $\sigma_b$ are the standard deviations. For both pairs of 
subsets $D > 4$, as well as for a mixture of two Gaussian distributions, $D>2$ 
is required for a clean separation between the modes. Additionally, we 
calculated distances between peaks and check if they exceed 
$3\sigma_{\rm tot}$ (where $\sigma_{\rm tot}$ comprises both statistical 
errors and measurement errors for a pair of subsets): 
$\Delta \mu_{\rm I-II} = 0.31$~kpc, whereas $3\sigma_{\rm I-II}=0.29$~kpc; 
and $\Delta \mu_{\rm II-III} = 0.44$~kpc, whereas 
$3\sigma_{\rm II-III}=0.42$~kpc. 

\begin{figure}
\vspace{2mm}
\centering
\includegraphics[width=8.9cm]{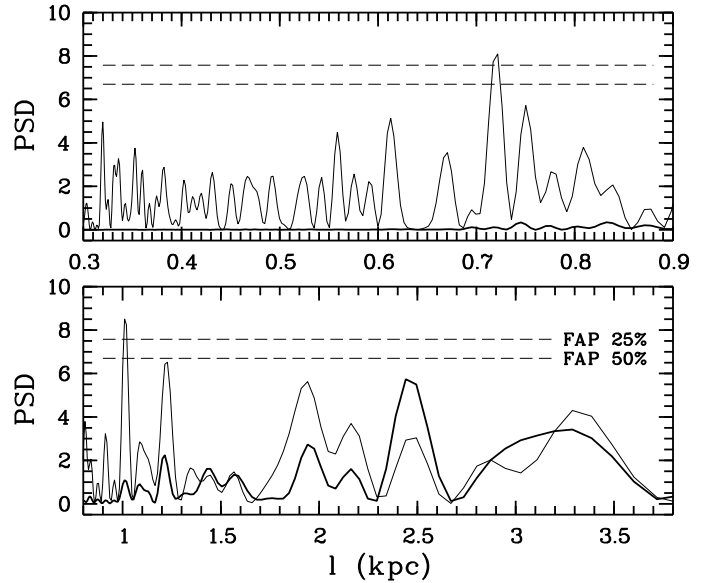}
\caption{Normalized power spectral density of the $U$ profile data from 
Fig.~\ref{fig:profile} (thick curves) and the function $p(s)$ (thin 
curves) for the ring. Dashed lines are the FAP levels of 25 and 50\%. 
See the text for more explanation.}
\label{fig:power}
\end{figure}

The most important result obtained from the analysis of the distribution 
over $l$ is the multiplicity of secondary peaks relative to the main one. 
The characteristic (mean) separations in subset II 
are about one and a half times greater ($1.46\pm0.25$, considering 
the internal positioning errors of pairs), while in subset III, they are 
about two times greater ($2.12\pm0.36$), and in subset IV, they are 
about three times greater ($2.88\pm0.49$) than the separation in subset I. 
We do not discuss subset IV as it contains only two 
measurements, which is insufficient for a statistic analysis.

Additionally, to estimate the power spectrum of our data, we computed 
the Lomb-Scargle periodograms for the $U$-band profiles along the ring shown 
in Fig.~\ref{fig:profile}, and for the function $p(s)$. The function 
$p(s)$ is a collection of Gaussians, centered at points of local maxima of 
brightness on the profiles, with $\sigma$ equal to the peak positioning error. 
The Gaussian amplitudes are assumed to be equal to the measured peak 
amplitudes, normalized to the maximum amplitude. If significant peaks were 
recorded in several bands ($U$, NUV, H$\alpha$), the largest normalized 
amplitude was adopted. Outside of the Gaussians, the function $p(s)=0$ is valid for all 
other points.

The periodograms, obtained using the technique developed by \citet{scargle1982}, 
\citet{horne1986}, and \citet{press1989}, also thoroughly explained by 
\citet{vp2018}, are presented in Fig.~\ref{fig:power}.
The Fourier analysis data, obtained for $U$ profiles and for functions 
$p(s)$, complement each other to support the presence of the spatial 
regularity of local maxima of brightness. The former shows characteristic 
low frequencies ($l>2$~kpc, bottom panel in Fig.~\ref{fig:power}), 
the latter represents high frequencies ($l<2$~kpc, upper panel).

The periodograms have noticeable peaks with a false-alarm probability $<25\%$ 
\citep[FAP; see][for details]{horne1986} of $l\approx0.72$ and $\approx1.0$~kpc, 
which corresponds to the characteristic separations in subsets I and II, 
respectively. Hence, the results of the Fourier analysis support the estimation of 
characteristic separations of local maxima of brightness in the ring based on 
their distributions in Figs.~\ref{fig:separate} and \ref{fig:hist} and also presented 
in Table~\ref{table:mean}.

\section{Discussion}

Earlier in this paper, we use NGC~628 as an example to discuss the effect of a shock 
wave on the formation of regularity in the distribution of young stellar groupings, 
along with a possible anti-correlation between shock wave signatures and the presence of 
chains of star complexes \citep{gusev2013}. This galaxy has two primary spiral 
arms, one of them showing an absence of shock wave and hosting the regular 
chain of star complexes, while the other does not. Nevertheless, both spiral arms 
in NGC~628 have the same regular characteristic separation between 
fainter star-formation regions that are adjacent to one another.

NGC~6217 is an example of a galaxy in which a similar regularity in the 
distribution of young stellar groups occurs in the ring located near the 
corotation radius. No shock waves are expected to be observed in this area. The detection 
of regularity here is a rather unusual result. In the corotation region, 
H\,{\sc i}/H$_2$ clouds, the progenitors of H\,{\sc ii} regions and young 
stellar groups are not expected to experience any significant mutual influence and 
they do not show a tendency towards concentration. A possible reason for the structural 
features observed in the ring of NGC~6217 may be that it may be, in fact, a pseudo-ring. 
A prominent bar can act as the driver of the Jeans instability wave. We note that the 
largest distances between adjacent maxima of brightness are observed directly in front of 
the bar (relative to the direction of rotation of the galaxy; see 
Fig.~\ref{fig:map}).

We obtained the characteristic distance, $\Lambda\sim700$~pc, which seems to correspond 
to a double wavelength of spacing regularity. \citet{elmegreen2018} in M100 
and \citet{gusev2013} in M74 (NGC~628) obtained the characteristic spacings 
of $\sim400$~pc, half as large. Our subset II corresponds to $\frc32\Lambda$, 
which may indicate the existence of a wavelength of $\frc12\Lambda$. 
The power spectral density of the $U$ profile has the largest peak at 
$l=2.45$~kpc (Fig.~\ref{fig:power}) that is close to $\frc72\Lambda$. Our linear 
resolution is insufficient to look for spatial regularities on the scale 
of $\frc12\Lambda$.

The characteristic spacing wavelength for hydrogen clouds is 
$\lambda_0 = 2c^2(G\sigma)^{-1}$, where $c$ is the sound speed and $\sigma$ 
is the mass column density of the gas \citep{elmegreen1983}. Spatial 
regularity in the ring of NGC~6217 indicates the constancy of parameters 
of the gas medium along the ring. The mass column density of H\,{\sc i} near 
the ring is $\sigma\approx5\Msun$pc$^{-2}$ \citep{driel1991}. The sound speed 
$c$ is a few km\,s$^{-1}$ for both $\lambda_0 = \Lambda$ and 
$\lambda_0 = \frc12\Lambda$, which is a typical speed in a diffuse H\,{\sc i} 
medium.

\section{Conclusions}

For the first time, a spatial regularity was found in the distribution of 
the young stellar population in the galaxy ring located near the 
corotation region. This refines our understanding of the physical 
processes that contribute to the formation of a regular wave of star formation 
in rings and arms. The presence or absence of shock waves does not appear to affect 
the genesis of regularity in regions of star formation.

The obtained characteristic separation, $\Lambda,$ between adjacent young stellar 
groupings in the ring of NGC~6217 is equal to $\approx670$~pc, that is twice as 
much as the characteristic separation obtained for young stellar groups in the 
spiral arms of M100 and M74. However, we suspect a characteristic separation 
with a scale of $\frc12\Lambda$ along the ring of NGC~6217, the same as was 
found earlier in spiral arms of M100 and M74. This may indicate a unified 
mechanism for the formation of such regularities in spiral arms and in the rings 
of galaxies. Achieving greater clarity around the characteristic separations in the 
rings requires the search and analysis of regularities in nearer ringed galaxies with 
improved spatial resolution.

\begin{acknowledgements}
We are grateful to the referee for their constructive comments. We would 
like to honor Prof. Yu.~N.~Efremov, who was the initiator of this 
research program, and who left us in August 2019. The authors acknowledge 
the use of the HyperLeda data base (http://leda.univ-lyon1.fr), the 
NASA/IPAC Extragalactic Database (http://ned.ipac.caltech.edu), 
Barbara~A.~Miculski Archive for space telescopes (https://archive.stsci.edu), 
and the IDL Astronomy User's Library (https://idlastro.gsfc.nasa.gov). 
This study was supported by the Russian Foundation for Basic Research 
(project no. 20-02-00080). A.G. acknowledges the support from the 
Lomonosov Moscow State University Development Program (Leading 
Scientific School ''Physics of stars, relativistic objects and galaxies'').
\end{acknowledgements}

\end{document}